\documentclass[prd,twocolumn,showpacs,tightenlines,%
superscriptaddress,nofootinbib,notitlepage,preprintnumbers]{revtex4}

\usepackage{hyperref}

\usepackage{amsmath}
\usepackage{amssymb}

\begin{document}

\title{Constraints on Shift-Symmetric 
Scalar-Tensor Theories with a Vainshtein Mechanism from Bounds on the 
Time Variation of $G$}

\author{Eugeny~Babichev} 
\affiliation{Laboratoire de Physique Th\'eorique d'Orsay,
B\^atiment 210, Universit\'e Paris-Sud 11,
F-91405 Orsay Cedex, France}
\affiliation{AstroParticule \& Cosmologie,
UMR 7164-CNRS, Universit\'e Denis Diderot-Paris 7,
CEA, Observatoire de Paris,
10 rue Alice Domon et L\'eonie
Duquet, F-75205 Paris Cedex 13, France}

\author{C\'edric~Deffayet} 
\affiliation{AstroParticule \& Cosmologie,
UMR 7164-CNRS, Universit\'e Denis Diderot-Paris 7,
CEA, Observatoire de Paris,
10 rue Alice Domon et L\'eonie
Duquet, F-75205 Paris Cedex 13, France}
\affiliation{${\mathcal{G}}{\mathbb{R}}
\varepsilon{\mathbb{C}}{\mathcal{O}}$, Institut d'Astrophysique
de Paris, UMR 7095-CNRS, Universit\'e Pierre et Marie
Curie-Paris 6, 98bis boulevard Arago, F-75014 Paris, France}

\author{Gilles~\surname{Esposito-Far\`ese}} 
\affiliation{${\mathcal{G}}{\mathbb{R}}
\varepsilon{\mathbb{C}}{\mathcal{O}}$, Institut d'Astrophysique
de Paris, UMR 7095-CNRS, Universit\'e Pierre et Marie
Curie-Paris 6, 98bis boulevard Arago, F-75014 Paris, France}


\begin{abstract}
We show that the current bounds on the time variation of the Newton constant $G$ can put severe constraints on many interesting scalar-tensor theories which possess a shift symmetry and a nonminimal matter-scalar coupling. This includes, in particular, Galileon-like models with a Vainshtein screening mechanism. We underline that this mechanism, if efficient to hide the effects of the scalar field at short distance and in the static approximation, can in general not alter the cosmological time evolution of the scalar field. This results in a locally measured time variation of $G$ which is too large when the matter-scalar coupling is of order one.
\end{abstract}

\date{July 7, 2011}

\maketitle

Many theories in which gravity is modified with respect to
general relativity (GR) contain, in addition to the metric, a
scalar field which is coupled directly to matter. Such
scalar-tensor theories appear naturally in low energy limits of
string theory and are also obtained from phenomenological
brane-world constructions (such as the DGP model
\cite{Dvali:2000hr}). Some are also of current interest as able
to produce an interesting cosmology via a large distance
modification of gravity. In such theories, in contrast to GR,
matter not only interacts via the helicity-2 graviton, but also
via the exchange of the scalar field. In general, one faces the
following dilemma: Either this field is coupled to matter with
gravitational strength, as required to produce order one
deviations from GR, but then the theory cannot pass local tests
of gravity, or the coupling is very small, but then there are no
significant effects of the scalar. A canonical example is the
Brans-Dicke theory~\cite{Jordan,Brans:1961sx} and its
extensions \cite{Damour:1992we} whose parameters are tightly
constrained by the local tests of gravity and observations of
binary pulsars (see for instance \cite{EspositoFarese:2009ta}).

A way out of this dilemma is provided by the Vainshtein
mechanism, first proposed in the context of massive gravity
\cite{Vainshtein:1972sx,Deffayet:2001uk} (a proof was recently
provided in \cite{Babichev:2009jt}). Indeed, close to localized
bodies, this allows to screen effects which lead to large
deviations from GR at large distances. This mechanism was also
shown to be present in the DGP brane model \cite{Dvali:2000hr} as
well as its decoupling limit \cite{Luty:2003vm}. It was later
generalized and shown to apply to a large class of scalar-tensor
models, called in \cite{Babichev:2009ee}``k-mouflage'' gravity
models, with a non-linear kinetic self-interaction of a scalar
field providing a self-screening of the scalar force {\it \`a la}
Vainshtein (hence the name k-mouflage). This class contains in
particular the Galileon model \cite{Nicolis:2008in}, and its
covariantized versions
\cite{Horndeski,Deffayet:2009wt,Deffayet:2011gz,Deser}. Many
applications of the Galileon model and its extensions to the
late-time acceleration, including minimally coupled
\cite{Deffayet:2010qz,Nesseris:2010pc,DeFelice:2010pv,%
DeFelice:2010as} as well as non-minimally coupled models
\cite{Chow:2009fm,Ali:2010gr,cosmologyN}, have been considered,
while various constraints coming from cosmology as well as from
local observations have been studied
\cite{Brax:2011sv,Kobayashi:2010wa,Kobayashi:2009wr,%
DeFelice:2010as,Nesseris:2010pc}.

In this paper, we point out that in spite of the fact that the
Vainshtein screening indeed allows to pass most of the
constraints coming from local observations by cutting off the
spatial variation of the scalar field near massive bodies, the
tests on the constancy of the Newton constant may easily rule out
many models. Indeed, we show that in many shift-symmetric models,
the evolution with time of the scalar field is (approximately)
the same everywhere and it follows its cosmological behavior. If
the scalar is directly coupled to matter, this induces a
variation of the Newton constant $G$, which is tightly
constrained by a number of observations (see, e.g., the review
\cite{Will:2005va}). The most stringent bounds come from
binary-pulsar data \cite{pulsar} and above all Lunar Laser
Ranging experiments \cite{LLR}, the latter giving $|\dot G/G| <
1.3\times 10^{-12} \,\text{yr}^{-1}$, or in terms of the Hubble
value today, $H_0$, $|\dot G/G| < 0.02 H_0$. As we will see
below, the time variation of the scalar field is generically of
order of the Hubble scale $H_0$ (unless it is in the
``cosmological'' screening regime with a tiny energy scale $M \ll
H_0$). This, whenever the direct coupling of the scalar field to
matter is of order of one, induces a too large variation of
Newton's constant, $|\dot G/G|\sim H_0$.

We consider the following general action,
\begin{equation}
\label{action}
\begin{aligned}
S = \frac{M^2_\text{P}}{2}\int d^4 x \sqrt{-g} \left( R +
\mathcal{L}_\text{s} + \mathcal{L}_\text{NL} \right)
+ S_m\left[\tilde{g}_{\mu\nu} ,\psi_m\right],
\end{aligned}
\end{equation}
where $R$ is the Ricci scalar of the metric $g_{\mu \nu}$,
$\mathcal{L}_{\rm s} = - \left(\partial\varphi\right)^2$ is the
standard kinetic term of a scalar field $\varphi$ (normalized to
be dimensionless), $\mathcal{L}_{\rm NL} $ describes some generic
nonlinear self-interaction of $\varphi$, and the matter fields
(collectively denoted as $\psi_m$) are minimally coupled to the
physical metric $ \tilde{g}_{\mu\nu} = \mathcal{A}^2(\varphi)
g_{\mu\nu}$. Because of this coupling, gravity is modified at
large distances through a scalar exchange, while GR is supposed
to be restored at small distances thanks the Vainshtein screening
effect made possible by the self-interaction $\mathcal{L}_{\rm
NL} $. This screening occurs for rather generic nonlinear
interaction terms $\mathcal{L}_{\rm NL} $ \cite{Babichev:2009ee}
and we do not specify a precise from of it. We however assume
that $\mathcal{L}_{\rm NL} $ is shift symmetric, i.e., it does
not change under the transformation $\varphi\to \varphi +
\text{const.}$ For example, we can choose $\mathcal{L}_\text{NL}$
to be in the Galileon or k-essence families. For dimensional
reasons $\mathcal{L}_{\rm NL} $ must contain a mass scale $M$. We
will assume that this is the only additional scale entering our
action. In general, this scale is fixed by phenomenological
requirements, e.g. to get present day acceleration of the
Universe from Galileons, $M$ should be of order of the Hubble
scale $H_0$.

Note that by a conformal transformation, action (\ref{action})
can always be rewritten in a form where matter is only minimally
coupled to one metric $\tilde{g}_{\mu \nu}$, going to the
so-called Jordan frame. Here we will work rather in the Einstein
frame (\ref{action}), with the understanding that our results
would apply to any theory whose action can be put in the form
(\ref{action}) by a suitable field redefinition. We will also not
consider effects that can arise, even in theories with minimal
coupling to gravity in the Einstein frame, due to the non-linear
kinetic coupling of the graviton to the scalar field (called ``kinetic
braiding'' in \cite{Deffayet:2010qz}).

The variation of Eq.~(\ref{action}) with respect to the metric
$g_{\mu\nu}$ gives the (modified) Einstein equations,
\begin{equation}
\label{Einstein}
M^2_\text{P} G_{\mu\nu} = T_{\mu\nu}^\text{(st)} +
T_{\mu\nu}^\text{(NL)} + T_{\mu\nu}^\text{(m)},
\end{equation}
where $T_{\mu\nu}^\text{(st)}$, $T_{\mu\nu}^\text{(NL)}$ and
$T_{\mu\nu}^\text{(m)}$ are respectively the energy-momentum
tensors for the standard scalar kinetic term, its nonlinear term,
and the matter contribution. The equation of motion for the
scalar field is
\begin{equation}
\label{eomp}
\nabla_\mu \left(\nabla^\mu \varphi + J^{\mu}_\text{NL} \right) =
-\alpha(\varphi)M^{-2}_\text{P} T^\text{(m)},
\end{equation}
where $\alpha(\varphi)\equiv
d\ln\left(\mathcal{A}\right)/d\varphi$, the nonlinear current
$J^\mu_\text{NL}$ is obtained by variation of
$\mathcal{L}_\text{NL}$ with respect to the gradient of the
scalar field, $J^{\mu}_\text{NL} \equiv - \frac12 \delta
\mathcal{L}_\text{NL}/\delta \varphi_{,\mu} $, and $T^\text{(m)}$
is the trace of the matter energy-momentum tensor in the Einstein
frame. Note that the field equations (\ref{eomp}) can always be
written as the divergence of a current, because of shift
symmetry. We also rescale the Planck mass so that
$\mathcal{A}(\varphi) =1$ at present. We stress that the trace of
the matter energy-momentum tensor defined in the Einstein frame,
$T_{\mu\nu}^\text{(m)}$, differs by the factor
$\mathcal{A}^{4}(\varphi)$ from the trace of the (conserved)
Jordan-frame energy momentum tensor. However, as we will see
below, the time variation of $\varphi$ is small (of order of the
Hubble scale or less), so that the change of $\varphi$ with time
can be neglected in the r.h.s. of (\ref{eomp}), giving only small
corrections. For instance, if $\mathcal{A}(\varphi) =e^\varphi$,
the approximation $\mathcal{A}^{4}(\varphi) \approx 1$ is valid
for $|\varphi |\ll 1$, i.e., for $|\Delta t| \ll H^{-1}$.

The cosmological evolution of the scalar field,
$\varphi_\text{cosm}(t)$ can be easily read from (\ref{eomp}),
\begin{equation}
\label{cosm1}
\ddot\varphi_\text{cosm} + 3H \dot\varphi_\text{cosm} - \nabla_0
\left(J^{0}_\text{NL}\right)= \alpha(\varphi)
M^{-2}_\text{P} T^\text{(m)}.
\end{equation}
Three different cosmological regimes can be identified. In a
regime where the cosmological energy density of the scalar field,
$\rho_\varphi$, is subdominant compared to the matter energy
density, $\rho_\text{m}$, $\rho_\varphi \ll \rho_\text{m}$, the
scalar field equation (\ref{cosm1}) is decoupled from the metric
equation (\ref{Einstein}). Then from the Einstein equations it
follows that $\rho_\text{m} = 3 M^2_\text{P} H^2$, thus the
r.h.s. of (\ref{cosm1}) is $\sim \alpha H^2$. If the scalar field
is away {}from the ``cosmological'' Vainshtein regime (i.e., when
the nonlinear term in the l.h.s. of (\ref{cosm1}) is negligible),
then from Eq.~(\ref{cosm1}) one can see that a particular
solution to (\ref{cosm1}) is $|\dot{\varphi}_\text{cosm}| \sim
\alpha H$. Notice that the general solution also contains a
homogeneous decaying solution $C_0 \exp\left(-3\int dt H \right)$
with an arbitrary constant $C_0$, however unless this constant
(or, equivalently, the initial condition) is fine tuned, the time
variation of $\varphi$ remains of order of $\alpha H$.

In the second regime matter is again dominant, $\rho_\varphi \ll
\rho_\text{m}$, but the scalar field is in the cosmological
Vainshtein regime. Then Eq.~(\ref{cosm1}) contains, in addition
to $H$, also the ``nonlinear'' scale $M$. Therefore the solution
to (\ref{cosm1}) generically contains a combination of scales $H$
and $M$. When this scale is small with respect to $H_0$, the time
evolution of $\varphi_\text{cosm}$ may thus be suppressed,
$|\dot\varphi_\text{cosm}|\ll H_0$. This is the cosmological
analog of the original Vainshtein mechanism.

The third regime is realized in the case when the scalar field is
dominant, $\rho_\varphi \gg \rho_\text{m}$, and in particular
when the late-time acceleration of the Universe is driven by the
scalar field. In this case, both the metric and the scalar field
equations depend only on one dimensionful parameter, $M$, which
is of order of $H_0$. We thus conclude that the typical value of
the present variation of the scalar field is the Hubble scale,
$|\dot\varphi_\text{cosm}| \sim H_0$.

Let us consider now the local effects caused by the conformal
coupling of the scalar field. For a slow time evolution,
$\varphi_\text{cosm}(t)$ can be written as the linear
approximation,
\begin{equation}
\label{cosm2}
\varphi_\text{cosm} (t) = \varphi_\text{cosm}(t_0) +
\dot\varphi_\text{cosm}(t_0)\, t.
\end{equation}
Note that (\ref{cosm2}) imposes the boundary value of the field
far from localized sources. The solution to the full equation of
motion (\ref{eomp}) at any point of space-time (including the
regions close to massive bodies, in particular, inside the
Vainshtein radius) depends on both time and space coordinates.
The key observation here is that thanks to the shift symmetry of
the equation of motion, the PDE (\ref{eomp}) allows separation of
variables in the following way,
\begin{equation}
\label{pansatz}
\varphi(t,r) = \varphi(r) + \dot\varphi_\text{cosm}(t_0) t +
\varphi_\text{cosm}(t_0),
\end{equation}
where $r$ is the distance to the source. It is not difficult to
see that the above ansatz (which has also been used in other
contexts \cite{Babichev:2010kj}) ``passes through'' the full
equation of motion (\ref{eomp}), giving an ordinary differential
equation of the second-order on $\varphi(r)$, with possible
remnants from the time-dependence in a form of a constant,
$\dot\varphi_\text{cosm}(t_0)$. The last two terms in the above
ansatz give the boundary condition for the PDE imposed by the
cosmological evolution (\ref{cosm2}), provided we choose the
radial dependent part of (\ref{cosm2}), $\varphi(r)$, to vanish
at infinity. Now, the ODE on $\varphi(r)$ is of the second order,
and supplied with two boundary conditions, $\varphi(r=\infty) =
0$ and $\varphi'(r=0) = 0$ (the last one comes from the
regularity at the origin), which is in general sufficient to find
a unique solution. Provided that this solution is non singular
(which is in some cases a strong mathematical requirement), and
assuming that (\ref{cosm2}) is a good approximation for the
time-dependent cosmological evolution of $\varphi$, our ansatz
gives a solution for all times to the field equation (\ref{eomp})
for a spherical source centered at $r=0$. The key point of this
paper is that the time derivative of $\varphi(t,r)$ is found to
be independent of $r$, i.e., it is set by the cosmological
evolution even inside the regions where the Vainshtein screening
operates.

It is worth mentioning that our ansatz (\ref{pansatz}) and
boundary conditions are in fact selecting a particular class of
solutions. Indeed, if formulated in terms of a Cauchy problem,
they implies $\varphi(t=0,r) = \varphi(r)$ and
$\dot\varphi(t=0,r) = \dot\varphi_\text{cosm}(t_0)$. In
principle, there is no reason not to choose some radially
dependent initial velocity, $\dot\varphi(r) = \mathcal{C}(r)$.
In contrast to (\ref{pansatz}), such solutions, however, are not
stationary, and they should relax to the stationary one, assuming
that the latter is stable. A numerical check of this relaxation,
as required by the nature of the field equations, goes however
far beyond the scope of this paper.

Let us consider a couple of  illustrative examples. First, as the non-linear term, we take one of the Galileon Lagrangians, namely, 
\begin{equation}
\label{example}
\mathcal{L}_\text{NL} = -M^{-2} \Box\varphi (\partial\varphi)^2,
\end{equation}
and the coupling to matter $\mathcal{A}(\varphi)= e^{\varphi}$.
It is not difficult to find that the evolution of $\varphi_\text{cosm}$ in two different cosmological regimes is in accord to our general findings
(we assume here a vanishing cosmological constant):
(i) when $\rho_\varphi \ll \rho_\text{m}$ and the nonlinear term is subdominant in (\ref{cosm1}) then 
$\dot\varphi_\text{cosm} = - 2 H$;
(ii) when $\rho_\varphi \ll \rho_\text{m}$ and the nonlinear term is dominant in (\ref{cosm1}) then 
$\dot\varphi^2_\text{cosm} = 2M^2/3$. The time dependent solution for the scalar field around a body of mass $m$ is then given by (\ref{pansatz}) with,
\begin{equation}
\label{solution2example}
\varphi'(r) = \frac{rM^2}{4}\left[-1+\sqrt{1+\frac{16 G m}{M^2r^3}\left(1+\frac{\dot\varphi_\text{cosm}^2}{2M^2}\right)}\right].
\end{equation}
Note that the last piece inside the parentheses is due to the cosmological time evolution of $\varphi$.
As a second example we consider a scalar field Lagrangian with the 
the signs of the scalar field kinetic terms flipped with respect to previous example,
and the coupling to matter $\mathcal{A}(\varphi)= e^{-\varphi}$. 
This Lagrangian allows self-accelerating solution \cite{Deffayet:2010qz}, with $H^2= M^2/(3\sqrt{6})$ and $\dot\varphi = \sqrt{6}H$,
while the time-dependent stationary solution for the scalar field is still given by (\ref{pansatz}) with the same radial-dependent part (\ref{solution2example}).

The time-dependence of $\varphi$ leads to a variation of the
effective Newton constant with time. This can be seen by making
the conformal transformation to the Jordan (physical) frame, with
the metric $ \tilde{g}_{\mu\nu}$. Generically, there are two
effects, which enter the final result for the evolution of
Newton's constant: the exchange of helicity-0 modes and the
rescaling of the coordinates via the conformal transformation of
the metric. It should be noted that in standard scalar-tensor
theories (without screening mechanisms), these effects are of the
same order, so that they even can compensate each others (as in
Barker's theory \cite{Barker}) giving no change of $G$. In our
case, however, the exchange of $\varphi$ is screened by the
Vainshtein mechanism, so that only one effect --- the stretching
of coordinates --- is important. As a result, the effective
Planck mass in the action gains a dependence on
$\varphi_\text{cosm}(t)$, $\tilde{M}_\text{P} =
\mathcal{A}^{-1}\left(\varphi_\text{cosm}\right) M_\text{P}$.
Thus, the observed Newton constant evolves with time as
\begin{equation}
\label{Gdot}
\big|\dot G/G\big|
\approx 2\alpha \dot\varphi_\text{cosm}(t).
\end{equation}
As we have seen before, depending on the regime,
$|\dot\varphi_\text{cosm}| \sim \alpha H$ (in the matter
domination regime) or $|\dot\varphi_\text{cosm} |\sim H$ (when
the scalar field is dominant). Therefore, presently one has
$|\dot G/G| \sim \alpha^2 H_0$ if the scalar field is subdominant
and away from the cosmological screening, and $|\dot G/G| \sim
\alpha H_0$ when the scalar field is dominant, in particular,
when it drives the late-time acceleration of the universe. This
applies in particular for a constant $\alpha$, i.e., a conformal
coupling $ \tilde{g}_{\mu\nu} = e^{2\alpha\varphi} g_{\mu\nu}$.

The observational constraints from Lunar Laser Ranging give
$|\dot G/G| < 0.02 H_0$ which is enough to rule out theories of
the kind considered here with a scalar coupling to matter of the
order of the gravitational one (i.e., $\alpha \approx 1$). In
order for a theory to explain the accelerated expansion of the
universe at present days, and pass the constraints on the
variation of $G$, one should assume $\alpha < 0.01$. It is
interesting to note that a similar constraint on the
matter-scalar coupling constant was obtained for the covariant
Galileon theory from a combined analysis of supernovae, baryonic
acoustic oscillations and cosmic microwave background
\cite{Ali:2010gr}.

Let us now briefly discuss the case of non shift-symmetric
theories. There is a class of such theories which can be put in
the form (\ref{action}) by suitable field redefinitions, in which
case our conclusions apply. When this is not the case, the
situation must be carefully reanalyzed. Indeed, the fact that the
ansatz (\ref{pansatz}) leads to a mere ODE to solve for
$\varphi(r)$ is a direct consequence of the shift symmetry and it
might be that some screening of the time variation of $G$ occurs
when this symmetry is lost. For example, we may introduce a mass
term in the action, $m^2\varphi^2$, which explicitly breaks the
shift symmetry. If the mass is big enough (say, much bigger than
the present Hubble scale, $H_0$), then the cosmological evolution
of $\varphi$ is suppressed, because the scalar follows the
minimum of the effective potential, $\dot\varphi_{min}\sim \alpha
H \dot H/m^2$. However, such theories do not possess either an
interesting self-accelerating scenario driven by the kinetic
term.

We also stress that even in the shift-symmetric case, it might be
that the ODE obeyed by $\varphi(r)$ does not possess regular
solutions, or that it leads to a stationary solution
(\ref{pansatz}) which happens to be unstable as a solution of the
PDE (\ref{eomp}). If so, it may open a way out of our
conclusions, necessitating the use of a more general ansatz than
(\ref{pansatz}) to solve the field equations, which could in turn
result in a Vainshtein suppression of the time variation of $G$
in the solar system.

It is also interesting to mention another possibility of avoiding
any significant time evolution of $G$, namely, to violate our
assumption of a conformal matter-scalar coupling. In particular,
in the relativistic MOND theory, called TeVeS \cite{TeVeS}, where
the physical metric is related to the Einstein one in a disformal
way, the time variation of the Newton constant is strictly zero
\cite{Bekenstein:2008pc}. This is a consequence of the different
scalings of time and space coordinates with
$\mathcal{A}(\varphi_\text{cosm})$ when one imposes that the
physical metric $\tilde g_{\mu\nu}$ tends to the Minkowski metric
at spatial infinity. This absence of time variation of $G$ also
applies to other theories with disformal coupling, in particular
to the improved relativistic MOND \cite{Babichev:2011kq}, where
the k-mouflage screening has been used to pass solar-system and
binary-pulsar constraints.

Let us finally underline that the DGP brane model, although
equivalent to a scalar-tensor theory of the Galileon type in the UV
(in particular in the decoupling limit \cite{Luty:2003vm}), is
not fully described by such a theory at cosmological scales (IR
limit). Therefore our analysis does not apply to the DGP model,
and this explains why Ref.~\cite{Lue:2002sw} did not find any
local time variation of $G$.

To conclude, we have shown that a generic scalar-tensor theory
with conformal coupling of a scalar field to matter, and with a
shift symmetry of the scalar Lagrangian, is tightly constrained
by the bounds on time variation of the Newton constant. The
models which fall into this category are not only standard (massless)
Brans-Dicke-like theories, but also those featuring a Vainshtein
screening mechanism due to the kinetic self-coupling, in
particular non-minimally coupled Galileon models. We argued that
the local time evolution of the scalar field is set by its
cosmological evolution and not screened by the Vainshtein effect,
which is only able to suppress the ``fifth force'' due to the
exchange of helicity-0 degree of freedom. The time derivative of
the scalar field is of order of the Hubble scale, unless the
whole universe is in the screening regime (``cosmological''
Vainshtein effect with $M \ll H_0$). This induces a time
evolution of Newton's constant of the same order. This result
applies for both the matter domination and the scalar field
domination cosmological regimes. It also does not depend on a
particular form of the non-linear self-interaction term, provided
it is shift-symmetric. The key point is that the time dependence
in (\ref{pansatz}), which is crucial in (\ref{Gdot}), is the same
irrespectively of the precise structure of
$\mathcal{L}_{\text{NL}}$. The evolution of the Newton constant
is, however, tightly constrained by observations, $|\dot G/G|
\lesssim 0.02 H_0$, therefore the conformal coupling on such
theories is constrained too. For example, if the non-minimal
coupling is of the form $\exp\left(2\alpha \varphi\right)$, then
$|\alpha|$ should be less than $10^{-2}$.


We thank J\'er\^ome Martin for discussions. The work of C.D. and
G.E.-F. was in part supported by the ANR grant ``THALES''.

\end{document}